\documentclass[twocolumn,aps,prl,showpacs]{revtex4}
\usepackage[T1]{fontenc}
\usepackage[latin1]{inputenc}
\usepackage{graphics}



\begin{document}

\title{Driving Rate Effects on Crackling Noise}

\author{Robert A. White}

\author{Karin A. Dahmen}

\affiliation{Loomis Laboratory, University of Illinois At Urbana, 1110 W. Green St., Urbana,
IL 61801, USA}

\date{File generated \today. Submitted for publication Feb. 6, 2002}

\begin{abstract}
Many systems respond to slowly changing external conditions with crackling noise,
created by avalanches or pulses of a broad range of sizes. Examples range from
Barkhausen Noise in magnets to earthquakes. Here we discuss how the scaling
behavior of the avalanche size and duration distribution and the power spectra
of this noise depend on the rate \( \Omega  \) at which the external conditions
are changed. We derive an exponent inequality as a criteria for the relevance
of adding a small driving rate \( \Omega  \) to the adiabatic model. We use
the zero temperature nonequilibrium random field Ising model to test our results,
which are expected to be applicable to a large class of systems with crackling
noise. They also agree with recent experiments on Barkhausen noise in various
materials.

\end{abstract}
\maketitle
Jerky response to smoothly varying forces has been enjoying increasing attention
for more than a decade. One reason for this is the surprising universality observed
in the noise statistics of vastly different systems. Earthquake faults, magnetic
tapes, vortices in type II superconductors, charge density waves, mechanically
failing materials, and many other systems respond to slowly changing external
conditions with discrete, impulse-like events (``avalanches'', or ``pulses'',
leading to familiar ``crackling noise'' \cite{SDM:01}) that span a huge range of
sizes: earthquakes for example are known to range from tiny tremors to devastating
magnitude 9 quakes on the Richter scale \cite{GutenRich}. Martensites crackle when
subjected to slow temperature or stress changes \cite{ORCGVMPLanes:95,CMOPVives:95}. Similarly, when a
magnetic tape is slowly magnetized by an external magnetic field, the process
happens in steps (``Barkhausen noise'' (BN)), by avalanches of re-orienting
magnetic domains, that range from microscopic to macroscopic in size \cite{BertBook}.
The avalanche size and duration distributions are often described by universal
power laws over a broad range of sizes. The scaling behavior on long length
scales is independent of the microscopic details, and only dependent on a few
basic properties, such as symmetries, dimensions, interaction range, and dynamics
(and a few others). In many systems this universality has been explained theoretically
as being due to either proximity to an underlying critical point \cite{SDKKRS:93},
or so called ``self-organized criticality'' (SOC) \cite{BakTangWies:98}.

Typically these systems are studied in the adiabatic limit where the rate \( \Omega  \)
at which the driving force \( F=\Omega t \) is increased, is taken to zero
allowing for a separation of the time scale of the fast local relaxation (given
by the avalanche durations) from the scale set by the slow external driving
(given by the rate \( \Omega  \)). For experiments and commercial applications
it is important to understand the effect of a finite but small driving rate
\( \Omega >0 \) on the pulse statistics. In this letter we study this question,
specifically focusing on the experimentally readily accessible BN in magnets
as the primary example. Our results are fairly general in nature and expected
to be applicable to a much larger class of systems with crackling noise.

A number of models have been used to study crackling noise ranging from interface
depinning models and sandpile models, to models with many interacting interfaces
due to domain nucleation. While we believe the main results we report in this
letter will apply to many approaches in the slow sweep rate regime we use a
variant of the zero temperature, nonequilibrium random field Ising model (ztneRFIM)
to illustrate our results. The ztneRFIM is a simple model for hysteresis and
avalanches (BN) in magnets \cite{SDKKRS:93,JRob:92}. Neglecting the microscopic details it
assumes a hypercubic lattice of spins $s_i = \pm 1$ with the Hamiltonian: 

\begin{equation} \mathcal{H}=-J\sum _{\left\langle i,j\right\rangle }s_{i}s_{j}- \sum _{i}(H(t)+h_{i})s_{i} \end{equation}
where the first term represents the ferromagnetically coupled nearest neighbors
(with coupling \( J>0 \)), \( H(t)=\Omega t \) is the external applied field,
and \( h_{i} \) is the random magnetic field with Gaussian distribution $\rho (h_{i})=\exp(-h_{i}^{2}/(2R^{2}))/(R\sqrt{2\pi} )$.
$h_{i}$ models uncorrelated quenched disorder. Other types of disorder, such
as random anisotropies, and random ferromagnetic bonds are expected to be in
the same universality class \cite{DaSe:96,VGOPlanes:95}.

We study the model at zero temperature, far from thermal equilibrium to model
systems in which the thermal relaxation time scale is much larger than experimental
time scales. One sweep in our simulation begins at \( H=-\infty \) causing all
spins to be down. We then raise the external field slowly and flip each spin
\( s_{i} \) when its local effective field 

\begin{equation} h_{i}^{eff}=H(t)+\sum _{\left\langle i\right\rangle }Js_{j}+h_{i} \end{equation}
changes from negative to positive. (Here $\left\langle i\right\rangle$ stands
for the nearest neighbors of site $i$.) A single spin flip takes a microscopic
time $\delta t \equiv 1$. It can trigger an avalanche of spin flips propagated
through the ferromagnetic interactions \cite{SDKKRS:93}. In the adiabatic limit $\Omega \rightarrow 0$,
the external magnetic field is kept constant during each avalanche, it is only
increased in between avalanches, until the next avalanche is triggered. The
avalanche time profile $V(t)$, which is the number of spins flipped during the
time interval of length $\delta t$ at time $t$ is proportional to the voltage
time profile $V(t)$ measured in the induction coil in a corresponding BN experiment
during a BN pulse. 

For the adiabatic case it has been shown \cite{SDKKRS:93} that near a critical amount
of disorder $R=R_c$ and a critical value of the magnetic field $H=H_c$ the avalanche
size and duration distributions scale on long length scales as $D_s(S)\sim S^{-\tau } {\cal{D}}_s(S h^{1/(\sigma \beta\delta)}, h/r^{\beta\delta})$
and $D_t(T)\sim T^{-\alpha} {\cal{D}}_t(T h^{\nu z/(\beta \delta)}, h/r^{\beta\delta})$,
respectively, where $r \equiv |(R-R_c)/R| $ and $ h \equiv |(H-H_c)/H|$. ${\cal{D}}_s$
and ${\cal{D}}_t$ are universal scaling functions and $\sigma$, $\nu$, $z$,
$\beta$, $\delta$, $\alpha$, $\tau$ are all universal exponents that are known
for the adiabatic case and are not all independent \cite{DaSe:96}. The scaling relation
between the size and duration, \( T\sim S^{\sigma \nu z} \) relates \( \alpha \)
to \( \tau \) through \( \alpha =\frac{\tau -1}{\sigma \nu z}+1 \). Numerical
simulations render \( \tau =1.60\pm 0.06 \) and \( \alpha =2.05\pm 0.05 \)
for this model \cite{PDS:99}. Away from the critical point the correlation length
$\xi \sim r^{-\nu} \Xi(h/r^{\beta\delta} )$ is finite, as is reflected by a
cutoff to the power law scaling regime at a largest avalanche $S^* \sim \xi^{1/\sigma \nu}$
and a largest duration $T^* \sim \xi^{z}$. (Here $\Xi$ is another universal
scaling function.)

At finite \( \Omega  \) new avalanches can be triggered by the increasing external
field before a previous one is completed. In this regime pulses observed in
the time series \( V(t) \) do not necessarily correspond to the avalanches
found in the adiabatic limit. (We stress here the subtle but necessary distinction
between {}``avalanche{}'' and {}``pulse{}''.) Since there is no mechanism
in the model for the sweep rate to effect the local dynamics of a spin flip
(a spin flips in one time step if its local field goes from negative to positive
regardless of how large the positive local field is) the observed differences
between pulses for finite \( \Omega  \), and avalanches for zero \( \Omega  \),
are a result of the simultaneous propagation of avalanches as well as the possible
internal changes to avalanches induced by spatial overlap of simultaneously
propagating avalanches at finite \( \Omega  \). For comparison with experiments
it is important to understand how the pulse size and duration distributions
are affected by temporal and spatial overlap of the avalanches as \( \Omega  \)
is increased from zero. 

A number of experimental results have been reported in the literature regarding
sweep rate effects on scaling exponents of pulse statistics\cite{BDM:94,DurZap:84,Zap:97}.
For BN in soft magnets without applied stress and with \( \alpha (\Omega =0)=2 \)
it has been reported that \( \alpha  \) and \( \tau  \) exhibit a linear sweep
rate dependence on \( \Omega  \) as \( \Omega  \) is increased from zero,
whereas experiments with applied stress give \( \alpha (\Omega =0)<2 \) and
at finite \( \Omega  \) the \( \alpha  \) and \( \tau  \) exponents do not
depend on \( \Omega  \). In the following we will show that these results can,
in fact, be generalized and explained by straight forward temporal superposition
of the adiabatic avalanches as \( \Omega  \) is increased.

There are two important threshold sweep rates, $\Omega_t$ and $\Omega_s$ that
mark the transitions to avalanches overlapping in time (\( \Omega >\Omega _{t} \)),
and in space {\it and} time (\( \Omega >\Omega _{s} \)):

\begin{figure}
{\par\centering \resizebox*{1\columnwidth}{!}{\rotatebox{270}{\includegraphics{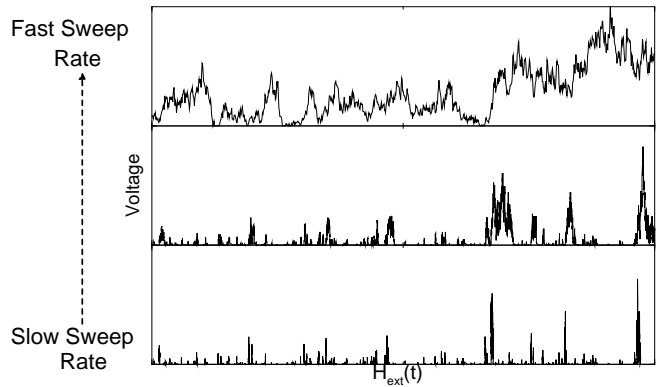}}} \par}

\caption{\label{timeseries}Simulated BN voltage $V$ (number of spins flipping per $ \delta t $)
as a function of applied external magnetic field $H_{\text ext}(t)$
(measured in units of coupling J), obtained using the ztneRFIM. At the bottom 
one can observe
separate pulses. As sweep rate is increased the pulses merge and become bigger
until no separate pulses can be observed.}
\end{figure}
(1) The lowest sweep rate $\Omega=\Omega_t$ at which on average successive avalanches
overlap in {\it time} so that one obtains a macroscopic uninterrupted pulse
(see Fig \ref{timeseries}), is called the upper bound of {}``slow{}''(note
that for $\Omega > \Omega_t$ the number of pulses is greatly reduced making
it difficult to obtain pulse statistics at all). To estimate $\Omega_t$ we set
the increase in $H$ during the mean avalanche duration \( \left\langle T\right\rangle  \)
equal to the mean separation \( \frac{1}{a(H)} \) (in \( H \)) between successive
avalanche seeds, and find \begin{equation} \label{omegat} \Omega _{t}\equiv \frac{1}{a(H)\left\langle T\right\rangle} = ( \frac{1}{\widetilde{a}(H) \left\langle T\right\rangle}) \frac{1}{L^{d}} \, , \end{equation}
where $a(H)\equiv \widetilde{a}(H)L^{d}$ is the non-universal number of avalanche
seeds per unit field increase, and \( \widetilde{a}(H) \) is independent of
the system size \( L^{d} \) (here \( L \) is the linear system size and \( d \)
is the dimension of the system). In general it is expected that $\Omega_t$ will
be proportional to $1/n$ where $n$ is the number of spins in principle available
to nucleate an avalanche (volume in our model, and surface for single domain
wall propagation). Eq.(\ref{omegat}) implies that in the thermodynamic limit
any finite sweep rate \( \Omega \) results in the loss of distinct pulses, since
all avalanches merge into only one big pulse.

(2) The lowest sweep rate \( \Omega _{s} \) at which successive avalanches tend
to overlap not only in {\it time}, but also {\it space} can be estimated as
follows: at $\Omega_s$ the expected number of new avalanche seeds triggered
by the increasing external magnetic field within the volume of a large propagating
avalanche of adiabatic size $S$, during its duration $T$, is a finite fraction
of $S$. If \( \tau >2 \) (i.e. \( \left\langle S\right\rangle  \) is not divergent)
\( \Omega _{s}\sim \frac{1}{T^{^{*}}}\gg \Omega _{t} \) (for large enough \( L \)).
If $\tau <2 $, in terms of the adiabatic quantities, we find \cite{WD:02}

\begin{equation} \label{omegas} \Omega _{s}\sim \frac{\xi ^{(\tau -2)/(\sigma\nu)}}{T^{^{*}}} \sim \xi ^{(\tau -2-\sigma \nu z)/(\sigma\nu)} \, . \end{equation}
For large enough system sizes $L$, for finite correlation length $\xi$, one
again obtains $\Omega_s \gg \Omega_t$. Therefore, the resulting time series
$V(t)$ in a BN experiment in the {}``slow{}'' sweep rate regime \( \Omega < \Omega _{t} \)
is simply a superposition of voltages produced by the individual simultaneously
propagating avalanches. 

Superimposing power law distributed pulses yields results consistent with the
experiments. The reason for this requires discussion: on the one hand, the time
between successive avalanche nucleations decreases for increasing sweep rate
$\Omega$. On the other hand the duration of the avalanche results from the dynamics
of the material and, in the slow sweep rate regime, does not depend on the \( \Omega  \).
This is true for all avalanche models we have considered. The resulting pulse
distributions are affected in two ways: (A) an avalanche of a certain size can
appear to take longer, since a second avalanche starts before the first one
ends, and runs longer (``swelling''). (B) Avalanches can completely disappear
(``absorption'') from the distributions due to their time interval of propagation
lying entirely within that of another avalanche. 

(A) To illustrate the relationship between swelling and the adiabatic exponents
we note that the largest avalanche nucleated during the propagation of another
avalanche of duration \( T \) (with \( \alpha >1 \)) is given by-\begin{equation}
\label{T_max}
T^{(2)}_{max}\sim (a(H)\, T\, \Omega )^{\frac{1}{\alpha -1}}
\end{equation}
 If \( \alpha >2 \) the avalanche of duration \( T \) will not swell appreciably
since \( T^{(2)}_{max}\sim T^{\varepsilon } \) where \( \epsilon <1 \) and
the new {}``swollen{}'' pulse is, at its largest, \( T+T^{(2)}_{max}\sim T \)
for large \( T \). For \( \alpha <2 \), however \( T^{(2)}_{max}>T \) is
no longer negligible: by iteration the process runs away for very small sweep
rate (for \( \Omega >\Omega _{t}\sim \frac{1}{L^{d+z(2-\alpha )}} \), see eqn(3)
for \( \alpha <2 \)). For \( \alpha =2 \) there is a linear increase in the
duration that is \( T\rightarrow \gamma T \) . In this case, as we see below,
absorption leads to a linear change of \( \alpha  \) with the sweep rate.

The discussion for the pulse size distribution runs analogously: The largest
secondary avalanche triggered by the increasing field during a primary avalanche
of duration T has size (for \( \tau >1 \))

\begin{equation}
\label{S_max}
S^{(2)}_{max}\sim \left( a(H)\, T\, \Omega \right) ^{\frac{1}{\tau -1}}
\end{equation}
 The sweep rate significantly affects the distribution only if \( S^{(2)}_{max}\geq S \).
This occurs if \( \frac{\sigma \nu z}{\tau -1}\geq 1 \), i.e. if \( \alpha \leq 2 \)
in agreement with the above argument for the duration distribution. If \( \alpha <2 \)
the effects of sweep rate on the size distribution can be seen in \cite{QueiBahi:01}
which we would argue results not in a decrease in the exponents, as suggested,
but in a decreasing region \( S<S^{*} \) over which the distribution can be
described by a power law at all (see \cite{WD:02} for more details) with \( S^{*}\sim \Omega ^{-\frac{(\alpha -1)}{(\tau -1)(2-\alpha )}} \).

(B) The relationship between absorption and the adiabatic exponents is similar.
As \( \Omega  \) is increased from zero, note, that for an avalanche to maintain
its identity (in terms of size and duration) it must not be absorbed into (i.e.
propagate simultaneously with) larger ones. We assume the avalanches are uncorrelated
in amplitude and have a nucleation density which can be described by a smoothly
varying function of h and the probability \( \mathcal{D}(T,\Omega ) \) that
a certain avalanche of duration T remains a pulse of duration T as \( \Omega  \)
is increased is given by

\begin{equation}
\label{finSweep}
\mathcal{D}(T,\Omega )=D(T,\Omega \rightarrow 0)\cdot \prod _{T'>T}(1-P(T',\Omega \rightarrow 0)),
\end{equation}
where \( P(T',\Omega \rightarrow 0) \) is the probability that the avalanche
of duration \( T \) is absorbed in an avalanche of duration \( T' \). A detailed
analysis of both swelling and absorption is given elsewhere \cite{WD:02}. One
finds that the effect of a slow but finite sweep rate depends on the adiabatic
value of the exponent $\alpha$: (1) For $\alpha(\Omega=0) > 2$, on long length
and time scales, we expect no change in the scaling exponents due to the introduction
of a slow sweep rate $\Omega<\Omega_t$. (2) For $\alpha(\Omega=0) = 2$ we find
a linear dependence of some of the exponents on the (slow) sweep rate: 

\begin{equation}
\label{alpha2}
\alpha (\Omega )=2-\kappa \Omega \; \; \; \; \; \; \; \; \; \tau (\Omega )=\frac{3}{2}-\frac{\kappa \Omega }{2}
\end{equation}
 where \( \kappa  \) is a non-universal constant. This is in agreement with
the results found earlier for the ABBM model \cite{ABBM:90t,ABBM:90e,DurBertFrac:95}, a single domain
wall model \cite{CizZap:97}, as well as for experiments in soft magnets with BN due
to the propagation of parallel domain walls \cite{DurZap:84}: all of these systems
have \( \alpha (\Omega =0)=2 \). (3) For $\alpha(\Omega=0) < 2$ the pulse perspective
is lost even for very small sweep rates since temporal overlap takes over (leading
to a large runaway pulse). For finite (but large) system size, in the slow sweep
rate regime $\Omega < \Omega_t $, so that one can actually still observe distinct
pulses, these pulses will be distributed according to the adiabatic scaling
forms, since in this case they tend to be very sparse on the time axis. This
result also agrees with experimental results of Durin and Zapperi on soft magnets
with stress induced anisotropies \cite{DurZap:84}.

\begin{figure}
{\par\centering \resizebox*{1\columnwidth}{!}{\rotatebox{270}{\includegraphics{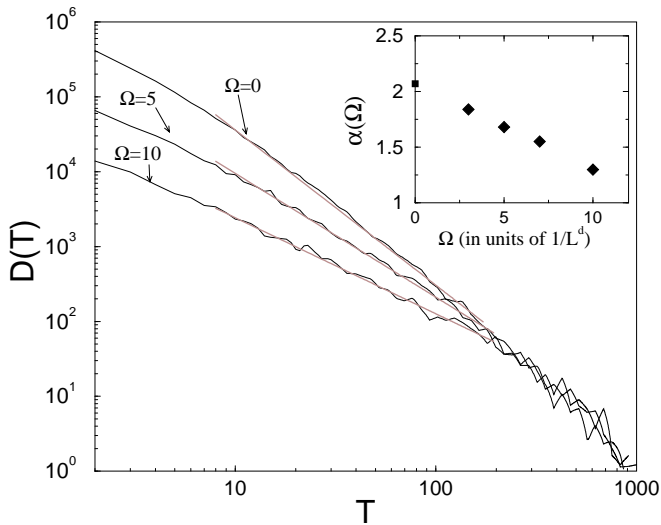}}} \par}

\caption{\label{a(h)}Distribution of avalanche durations $T$ (number of time steps $\delta
t=1$) at $H\sim H_c$ and $R\sim R_c$ for different sweep rates for the
ztneRFIM in \protect\( d=3\protect \) dimensions. Inset shows linear sweep
rate dependence of the exponent \protect\( \alpha \protect \) taken from power
law fits to the duration distributions. The simulated systems have \protect\( L^{d}=500^{3}\protect \)
spins. The data has been averaged over \protect\( 20\protect \) disorder realizations.}
\end{figure}

Numerically in the 3-d ztneRFIM we find that \( \alpha (\Omega ) \) appears
to have a linear dependence on \( \Omega \) since the ztneRFIM has \( \alpha =2.05\pm .05 \),
i.e. very close to 2 (see fig. \ref{a(h)}). We would expect, however, that
if \( \alpha \)>2, for very large durations (roughly \( T\geq 10^{\frac{1}{\varepsilon }} \)
where \( \alpha =2+\varepsilon  \) , for our problem \( \epsilon =.05 \))
we would find, in the large system size limit, the finite but slow sweep rate
 to leave the exponent \( \alpha  \) unchanged. We estimate the system size required
to bring the exponent to within the error bars (i.e. \( \pm .05\))  of the adiabatic value for \(\alpha = 2.05\) and
we find the system size must be orders of magnitude larger than the largest system we can simulate (\( 500^{3} \)) (see \cite{KPDRS:99} for details of the base algorithm).  It should be easier to observe the exponent independence
in models with \( \alpha \) significantly greater than 2. 

Conveniently it can be shown \cite{TWD:01} that the high frequency scaling behavior
of the \textit{power spectra} (PS) of the the BN voltage time series is independent
of the sweep rate for \( \Omega <\Omega _{s} \). This is a direct consequence
of the time series being a mere superposition of the uncorrelated adiabatic
avalanches. For Poissonian distributed avalanche nucleation events (as in the
ztneRFIM) the PS of the time series is a simple superposition of the PS of the
individual pulses \cite{TWD:01,Mazz:62} and we note no difference in the scaling even
in the lowest frequency of the PS. Since we are able to directly extract many
of the exponents associated with the adiabatic avalanches (z,\( \nu  \)z,\( \sigma \nu z \),\( \frac{3-\tau }{\sigma \nu z} \),\( \frac{2-\tau }{\sigma \nu z} \))
from the PS regardless of sweep rate, collapses of the PS provide a useful tool
for probing the collective behavior without needing to address temporal overlap
of avalanches. For \( \Omega >\Omega _{t} \) where pulse statistics are lost
and avalanche interactions become important as seen in \cite{HwaKardar:92} one \textit{must}
use spectral tools.

In summary we have found that the experimental observations of sweep rate dependence/independence
of pulse duration and size distribution exponents can be understood as a consequence
of temporal superposition of the avalanches that would be observed in the adiabatic
limit. This conclusion is borne out by our numerical work on the ztneRFIM where
we know that the pulses are simply a superposition of the adiabatic avalanches.
We also provide criteria for determining for which sweep rates pulse statistics
provide insight into the adiabatic collective behavior and when spectral methods
will serve better. 

\begin{acknowledgments}

We thank Jim Sethna and Mike Weissman for very clarifying discussions. This
work was supported by the Materials Computation Center, grant NSF-DMR 99-76550,
and NSF grant DMR 00-72783. K.D. also gratefully acknowledges support from an
A.P. Sloan fellowship. We thank IBM for a very generous equipment award that
made extensive numerical simulations for this work possible.

\end{acknowledgments}

\bibliography{RFIM}
\end{document}